
\magnification1200

\font\bfuno=cmbx12

\font\sf=cmss10
\font\title=cmss17

\def\V{{\cal V}}
\def\C{{\cal C}} 
\def\D{{\cal D}}
\def\A{{\cal A}}
\def\M{{\cal M}}
\def\sss{\scriptscriptstyle}

\input amssym.def
\input amssym.tex


\def\mathbb{\Bbb}
\def\R{{\Bbb R}}
\def\Rfi{\R}

\def\Rn{{\R^n}}

\def\Dif{\,\hbox{\sf D}}
\def\Rn{{\R^n}}
\def\pinf{{p_{\sss\infty}}}
\def\ainf{{a_{\sss\infty}}}

\def\dom{{\hbox{\rm dom}\,}}

\mathchardef\emptyset="083F

\hsize17.5truecm
\vsize26.2truecm
\hoffset-.8truecm
\voffset-1truecm

\noindent This survey appeared  in: Nonlinear Evolution Equations and Dynamical Systems, {\it Proceedings of 
the Vth NEEDS Workshop}, 
Springer-Verlag, 173--180 (1990).
\bigskip\bigskip

\noindent
\title Liouville-Arnold Integrability 

\smallskip

\noindent
for Scattering under Cone Potentials \rm

\medskip

\noindent
\sl Gianluca Gorni\/$^1$ and Gaetano
Zampieri\/$^2$ \rm

\medskip

\noindent
\halign{$#$\ &#\hfil\cr
^1&Dipartimento di Matematica e Informatica,
Universit\`a di Udine,\cr
&Via Zanon 6, I--33100 Udine UD, Italy\cr
^2&Dipartimento di Matematica Pura e Applicata,
Universit\`a di Padova,\cr
&Via Belzoni 7, I--35131 Padova PD, Italy\cr}

\bigskip

\centerline{{\bfuno 1. Introduction}}

\medskip

The problem of scattering of particles on the line 
with repulsive interactions, gives rise to some
well-known integrable Hamiltonian systems, for
example, the nonperiodic Toda lattice or
Calogero's system.  The aim of this note is to
outline our researches which proved the
integrability of a much larger class of systems,
including some that had never been considered,
such as the scattering with very-long-range
interaction potential. The integrability of all
these systems survives any small enough
perturbation of the potential in an arbitrary
compact set. Our framework is  based on the concept
of cone potentials, as  defined below, which
include the scattering on the line as a particular
case. 

In Section 2 we present some remarkable examples
 that are covered by our
theory. In Section 3 we discuss the
related literature. Finally, in Section 4 we write
down the statements of the results.
\smallskip 

Let us consider a particle
$q\in\Rn$ that moves in the field generated by a
potential~$\V$: $$\dot q=p\,,\qquad
  \dot p=-\nabla\V(q)\,.  
  \eqno(1.1)$$
Suppose that $\V\in C^2(\Rn)$ and that

\smallskip

\item{i)} \sl $\V$ is bounded from below;\rm

\item{ii)} \sl the force $-\nabla\V(q)$ always
lies in a closed, convex cone~$\bar\C$ which is
``proper'', i.e., containing no straight lines
({\it``cone potential''}). \rm

\smallskip

\noindent
It is then well-known and easy to prove that
all trajectories have a finite {\it asymptotic
velocity} $\pinf:=\lim_{t\to+\infty}p(t)$, as we
are going to see.

Observe first that the system admits the first
integral of energy ${1\over2}|p(t)|^2+\V(q(t))$.
From~i) we see that every trajectory has a
bounded speed, so that it is also globally
defined in time. 

On the other hand, property ii) is equivalent to
the existence of a  basis
$\{b_1,\ldots,b_n\}$ of~$\Rn$ such that
$-\nabla\V(q)\cdot b_i\ge0$ for all~$q\in\Rn$
and all~$i$ (see e.g. Gorni, Zampieri [GZ1]).
Hence for every trajectory $(p(t),q(t))$ the map
$t\mapsto p(t)\cdot b_i$ is a monotone function
for all~$b_i$. Monotone bounded functions defined
on~$\R$ always admit a finite limit
at~$+\infty$, whence the claim is proved.

The existence of asymptotic velocities is
a characteristic of scattering systems. Cone
potential systems have actually been introduced
as a generalization of the problem of
a system of mutually repelling particles in one
dimension.

The $n$ components of~$\pinf$ are trivially
constants of motion. Some heuristics make them
also likely candidates for being~$n$
independent first integrals in involution as
called for by the theory of Integrable
Hamiltonian Systems (see e.g. Arnold [A]). The
conjecture, suggested by Gutkin in~[Gu1], is
false  without further restrictions on the
potential, though. In particular, the asymptotic
velocity may not even be a continuous function
of the initial data (see~[GZ1],
Section~3), and sometimes there are geometrical
obstructions to the global independence of its
components, even if they happened to be smooth
(see~[GZ3], Sections~1 and~5).

In the papers [GZ1,2,3] the authors develop a
theory of the $C^k$ ($2\le k\le+\infty$)
regularity of the asymptotic velocity and
Liouville-Arnold integrability for cone
potentials, whose main points we are going to
sketch here.

Let $\C$ be the convex cone generated by the
forces, that is, the smallest convex cone
that contains all vectors $-\nabla\V(q)$ for
$q$~in the domain of~$\V$ (the potentials
in the examples below come with their
natural domains; for simplicity think of
the whole of~$\Rn$ for now). Let us denote by~$\M$
the subset of~$\Rn\times\Rn$ of the initial data
(velocity$,{}$position) that are {\it
asymptotically regular}, i.e., that lead to an
asymptotic velocity with {\it positive} scalar
product with all vectors of the closure of~$\C$:
$$\pinf\cdot v>0\qquad\forall v\in\bar\C
  \eqno(1.2)$$
(in the language of convex sets, this amounts to
saying that $\pinf$ lies in the {\it interior}
of the {\it dual cone} $\C^*$).
Given these basic definitions, the theory splits
into two distinct parts:

\item{I.} $\M$ is clearly an {\it invariant}
subset of the phase space. Under very general
and natural hypo\-theses on the potential (made
explicit in~[GZ3]), $\M$ is also {\it open} and
{\it nonempty}. The bulk of~[GZ1] is devoted to
proving that the asymptotic velocity~$\pinf$ is
a smooth ($C^k$, $2\le k\le+\infty$) function of
the initial data on~$\M$, if suitable
additional hypo\-theses are made on~$\V$.
The starting point is the integral expression 
for the asymptotic velocity:
$\pinf=p(0)-\int_0^{+\infty}\nabla\V(q(t))dt$.
Two basic ways are found to apply to this formula
the usual theorems of continuity and
differentiability of integrals depending on
parameters. If $\V$ decays {\it exponentially}
at infinity (see a later section for precise
statements), we give Gronwall-like estimates on
the growth of the derivatives of~$q$ with
respect to the initial data. If instead $\V$ is
assumed {\it convex}, then a simple
{\it monotonicity} condition on the Hessian
matrix of~$\V$ enables us to build a Liapunov
function that checks the growth of the
derivatives of~$q$ without special decay
requirements on~$\V$.
In both settings it is proved, again in~[GZ1],
that the components of~$\pinf$, in addition to
being differentiable, are independent and in
involution on~$\M$, and the range of~$\pinf$
(on~$\M$) is exactly the interior of the dual
cone~$\C^*$. In particular, the Hamiltonian
system is integrable if it is restricted to the
open, nonempty, invariant set~$\M$.

\goodbreak

\item{II.} In [GZ1,3] we provide examples where
$\M$ does not coincide with the whole phase
space $\Rn\times\Rn$. This can happen either for
a trivial reason such as the presence of an
equilibrium ($\nabla\V(q)=0$), or for subtler
reasons. In those same papers we present three
different hypotheses on the potential~$\V$ that
guarantee that all initial data are
asymptotically regular.  Two of
them have close parallels in Gutkin [Gu2]
and Hubacher [Hu].

\noindent
If we collect together the hypotheses of~I with
those of~II, we come up with wide classes of {\it
completely integrable systems}.

\goodbreak

The paper [GZ2] shows that if we restrict to
potentials with fast enough decay at infinity,
then the motion of the particle $q$ is
asymptotically rectilinear uniform: $q(t)=\ainf+
\pinf t+o(1)$ as~$t\to+\infty$, and the
{\it asymptotic data} $(\pinf,\ainf)$, as
functions of the initial data, define a
{\it global canonical diffeomorphism
(``asymptotic map'')} $\A\colon(p,q)
\mapsto(P,Q)$, which brings the original
Hamiltonian system~(1.1) into the {\it normal
form}: 
$$\dot P=0\,,\qquad \dot Q=P\,.
  \eqno(1.3)$$

A noteworthy property of the systems
introduced in [GZ1,2,3] is that the
integrability,
 and (when it
is the case) the reduction to normal form via
asymptotic data, endure any small enough
perturbation of the potential~$\V$ in any
arbitrary compact set of~$\Rn$. Such ``structural
stability'' property seems to be unusual in the
literature on integrable Hamiltonian systems,
that have been studied mostly by algebraic
techniques which are rather ``rigid''.

\bigskip\vfill\eject

\centerline{{\bfuno 2. Examples}}

\medskip

Some remarkable examples of cone potential
systems that are covered by our theory of 
$C^\infty$-integrability are the following.
Here $N\ge1$ and the vectors  $v_\alpha\in\Rn$,
with $\alpha=1,\ldots,N$, generate a proper cone
(i.e., there exists $\bar v\in\Rn$ such that
$\bar v\cdot v_\alpha>0\;\; \forall\alpha$).

\smallskip

\item{a)} Toda-like, or exponential potentials
$$\V(q):=\sum_{\alpha=1}^N c_\alpha 
  e^{-q\cdot v_\alpha}\,,\qquad
  c_{\alpha}>0,\quad q\in \Rfi^n\,.
  \eqno(2.1)$$

\item{b)} Finite sum of inverse powers:
$$\V(q):=\sum_{\alpha=1}^N
  {1\over(q\cdot v_\alpha)^{r_{\alpha}}}\,,\qquad
  q\in \{\bar q\in\Rfi^n\;\colon\;
  \bar q\cdot v_\alpha>0\;\forall\alpha\}\,,
  \eqno(2.2)$$
in three different hypotheses:

\medskip

\itemitem{b1)} $r_\alpha>0$ arbitrary but the
vectors satisfy $v_\alpha\cdot v_\beta\ge0$;

\itemitem{b2)} $v_\alpha$ arbitrary but all the
exponents satisfy $r_\alpha>1$;

\itemitem{b3)} $v_\alpha$ arbitrary but the
exponents $r_\alpha$ are all equal to an~$r>0$.

\medskip

\item{}
In particular b3) includes the scattering of
particles in  one dimension with inverse
$r$-power potential, for arbitrary $r>0$:
$$\V(q):=\sum_{1\le i<j\le n}
  {1\over(q_i -q_j)^r}\,,\qquad q\in
  \{(q_1,\ldots,q_n)\in\Rfi^n\,:\,
  q_1<q_2<\ldots<q_n \}\,.
  \eqno(2.3)$$
The case $r=1$ is the Coulombian potential,
whilst $r=2$ is the Calogero potential. The
exponent $r=1$ divides what are usually called
short range potentials ($r>1$) from the long
range ones ($0<r\le1$). 

\medskip

\item{c)} Even longer range potentials are the
inverse-logarithmic type:
$$\V(q):=\sum_{\alpha=1}^N
  {1\over(\ln(1+q\cdot v_\alpha))^r}\,,\qquad
  q\in \{\bar q\in\Rfi^n\;\colon\;
  \bar q\cdot v_\alpha>0\;\forall\alpha\}\,,
  \eqno(2.4)$$
for arbitrary $r>0$.

\medbreak

For potentials a), b1) with $r_\alpha>1$, and b2),
there are both asymptotic velocities and phases,
with consequent reduction to normal form. In
case~b3) with $0<r\le1$ and c) we   prove
integrability through asymptotic velocities but
we do not have the  reduction to normal form
through asymptotic map.

All the previous examples share the ``finite
sum'' form, that gives rise to {\it polyhedral}
cones~$\bar\C$ and $\C^*$. This is not an
essential feature of the theory, as we show
in~[GZ2], where we prove the
$C^\infty$-integrability of the system in three
dimensions whose potential is 
$$\V(x,y,z):={2z\over z^2-x^2-y^2}\exp
  \Bigl(-{z^2-x^2-y^2\over2z}\Bigr)
  \eqno(2.5)$$
defined on the set $\D^\circ=
\{(x,y,z)\in\Rfi^3\;:\;z>\sqrt{x^2+y^2}\}$. Here
the cone~$\C$ and the interior of the dual
turn out to coincide with the domain of~$\V$,
which is a {\it circular} cone.

\goodbreak
\bigskip

\centerline{{\bfuno 3. Related papers}}
 
\medskip

Special instances of integrable systems with
cone potentials, such as the non-periodic Toda
lattice and Calogero's system, have been
well-known since the seventies. The
techniques of integration did not make any use of
the asymptotic velocity (see e.g. Moser~[M]).

Gutkin in~[Gu1] introduced the concept of cone
potentials with the conjecture that the
integrability of those special systems could be
derived from a general theory.
Stimulated by Gutkin's idea, Oliva and Castilla
in~[OC] gave a rigorous proof of
$C^\infty$-integrability via asymptotic
velocities for a class of finite-sum potentials
which decay exponentially using
Dynamical Systems techniques, very different
from our approach.

The results of Hubacher's paper~[Hu] 
overlap partially with the ones of~[GZ3]. It is
concerned with systems of mutually repulsive
particles on the line, that fall under what we
call ``finite-sum form'' potential, as in
formula~(2.3).  Hubacher states the integrability
through asymptotic data  in the case of ``short
range'' potentials (roughly corresponding to $r>1$
in~(2.3)) by invoking the results of
Simon~[S] and Herbst~[He]. Those authors had
studied the scattering of $n$ mutually repulsive
particles in~$\Rn$, through an approach totally
different from ours. The proof of smoothness
was obtained there by solving a kind of Cauchy
problem at infinity with the asymptotic data
playing the role of initial data. The method is
remarkably simple for short range potentials, but
it gets into complications for long range
potentials ($0<r\le1$), and does not seem to
cover for example the inverse-logarithm
case~(2.4). (Incidentally, Hubacher also obtains
another interesting result outside the spirit of
the present paper). 

Moauro, Negrini and Oliva [MNO]  recently gave
a proof of {\it analytic} integrability
for the systems of cases~a) and~b) in Section 2
by means of Dynamical Systems 
techniques. 

Finally, we remark that the problem of the mere
existence of asymptotic velocities and phases for
scattering systems of mutually repelling
particles in~$\Rn$ was studied by various authors
in the sixties and seventies, see for instance
[Ga] and the references contained therein.

\goodbreak
\bigskip

\centerline{{\bfuno 4. Statements of Results}}

\medskip

Given a smooth function $\V\colon\Rn\to\R$,
we will denote by~$\nabla\V$ its gradient, as a
column vector, and $\Dif^m\V$~will be its $m$-th
differential, regarded as a multilinear map from
$(\Rn)^{m-1}$ into~$\Rn$, endowed with the
norm
$$\|\Dif^m\V(q)\|:=\sup\{
  |\Dif^m\V(q)(x^{(1)},\ldots,x^{(m-1)})|\;:\;
  x^{(i)}\in\Rfi^n,\;|x^{(i)}|\le1\}.
  \eqno(4.1)$$

In the sequel, $\C$ will be the convex cone
generated by the force $-\nabla\V$ of a cone
potential, and $\D$ will be the dual of~$\C$:
$$\D:=\bigl\{v\in \Rfi^n\, :\,
   w\cdot v \ge 0\,
   \quad \forall
   w\in \C\bigr\}\,.
  \eqno(4.2)$$
Throughout this section the function $\V$ can be
assumed to be defined either on all of~$\Rn$ or
on a set of the form~$q+\D^\circ$. 
In addition to the asymptotic velocity, we are
also interested in the existence and the
smoothness of the {\it asymptotic map}
$${p_{\sss\infty}(\bar p,\bar q)\choose
  a_{\sss\infty}(\bar p,\bar q)}=
  \A(\bar p,\bar q):=\lim_{t\to+\infty}
  \A_t(\bar p,\bar q)\,.
  \eqno(4.3)$$

\goodbreak

We start with the assumptions corresponding to
point~I of the Introduction, that is, those
leading to integrability on the set~$\M$.

\bigbreak

\strut\hbox{\bf Hypothesis 4.1 } \sl There exists
$q_{\sss0}\in\dom\V$ and a nonnegative, weakly
decreasing and in\-te\-grable function
$h_{\sss0}\colon[0,+\infty[\to\R$ such that \rm
$$q\in q_{\sss0}+\D\quad\Rightarrow\quad
  |\nabla\V(q)|\le h_{\sss0}\bigl(
  \hbox{{\rm dist}}
  (q,q_{\sss0}+\partial\D)\bigr)\,.
  \eqno(4.4)$$

\bigbreak

\strut\hbox{\bf Hypotheses 4.2 } \sl The
potential $\V$ is a $C^{m+1}$, $m\ge2$, function.
For all $1\le i\le m$ there exist $q_i\in\Rn$,
$A_i\ge0$, $\lambda_i>0$ such
that  
$$q\in q_i+\D\quad\Rightarrow\quad
  \|\Dif^{i+1}\V(q)\;\|\le
  A_i\exp\Bigl(-\lambda_i\,
  \hbox{{\rm dist}}\bigl(
  q,q_i+\partial\D\bigr)
  \,\Bigr)\,.
  \eqno(4.5)$$
\rm

\bigbreak

\strut\hbox{\bf Hypotheses 4.3 } \sl The
potential $\V$ is a $C^{m+1}$, $m\ge2$, function.
For all $1\le i\le m$ there exist $q_i\in\Rn$,
and a weakly decreasing function
$h_i\colon[0,+\infty[\to\R$ such that  
\item{1)} $\V$ is convex on
  $q_{\sss1}+\D$; 
\item{2)} for all
  $q^\prime,\;q^{\prime
  \prime}\in q_{\sss1}+\D$ and all $z\in\Rn$
  we have
  $$q^{\prime\prime}\in q^\prime+\D\quad
    \Rightarrow\quad
    \Dif^2\V(q^{\prime\prime})z\cdot z\le
    \Dif^2\V(q^\prime)
    z\cdot z\hbox{;}
    \eqno(4.6)$$
\item{3)} for all $i$,
  $\int_0^{+\infty}\mskip-7mu 
  x^i\, h_i(x)dx<+\infty$ and
  $$q\in q_i+\D\quad\Rightarrow\quad
    \|\Dif^{i+1}\V(q)\|\le h_i\Bigl(\,
    \hbox{{\rm dist}}\bigl(q,q_i+
    \partial\D\bigr)\,\Bigr).
    \eqno(4.7)$$

\rm

\goodbreak
\bigskip

Here is the precise statement of the
integrability on the set~$\M$.

\bigskip

\strut\hbox{\bf Proposition 4.4 } \sl Assume
that $\V$ is a cone potential and Hypotheses~4.1
and either~4.2 or~4.3. Then the components of
the asymptotic velocity~$\pinf$ are $C^m$ first
integrals, independent and in involution on the
set~$\M$ of the asymptotically regular initial
data. If, moreover, the functions $h_i$ of
Hypotheses~4.1 and (when the case)~4.3 verify
$\int_0^{+\infty} x^{i+1}\,h_i(x)\,dx<+\infty$
for all $0\le i\le m+1$, then the asymptotic
map~$\A$ exists on~$\M$, it is a global canonical
$C^m$ diffeomorphism from~$\M$ onto
$\D^\circ\times\Rn$, and it brings the restricted
phase space system
$$\dot p=-\nabla\V(q)\,,\quad
  \dot q=p\,,\qquad
  p,q\in\M
  \eqno(4.8)$$
into the normal form \rm
$$\dot P=0\,,\quad\dot Q=P\,,\qquad
  (P,Q)\in\D^\circ\times\Rn\,.
  \eqno(4.9)$$

\goodbreak
\bigskip

Next we present the assumptions under which we
can prove that the set~$\M$ coincides with the
whole phase space. The first one guarantees that
all trajectories are contained in some
translation of~$\D$, so that the asymptotic
velocity lies always in the closed set~$\D$,
although possibly on the boundary.

\bigskip

\strut\hbox{\bf Hypotheses 4.5 } \sl For each
$E>0$  there exists a $q_{\sss E}\in\Rn$ such
that \rm 
$$q\in \Rfi^n\backslash
  (q_{\sss E}+\D)
  \quad\Rightarrow\quad 
  \V(q)\ge E\,.
  \eqno(4.10)$$

\bigbreak

The following hypothesis is not so transparent at
first view, but one of its points is that the
cone of the forces~$\C$ should be of width not
larger than~$\pi/2$. No finite-sum form of the
potential is called for, though. The circular
cone example of the previous Section falls into
this category.

\bigbreak

\strut\hbox{\bf Hypotheses 4.6 } \sl For each 
$q^\prime\,,\,q^{\prime\prime}\in\Rn$ 
such that 
$q^{\prime\prime}\in  q^\prime+\D$,
and for each
$v\in\bar\C\backslash\{0\}$ 
there exists 
$\varepsilon>0$ such that \rm
$$\Bigl(\; q\in q^\prime+\D
  \quad\hbox{and}\quad 
  q\cdot v\le
  q^{\prime\prime}\cdot v \;\Bigr)
  \quad\Rightarrow\quad
  -\nabla\V(q)\cdot v\ge\varepsilon\,.
  \eqno(4.11)$$

\bigbreak

The last assumptions
concern the potentials which can be
written as finite sums of one-dimentional
functions (``finite-sum potentials''):
$$\V(q):=\sum_{\alpha=1}^N
  f_\alpha(q\cdot v_\alpha)\,,
  \eqno(4.12)$$
where $v_1,\ldots,v_N$ are given nonzero vectors
in~$\Rn$ ($N\ge1$, no relation to $n$), and the
functions $f_1,\ldots,f_N$ are real functions of
one variable, whose domains are each
either $\R$ or the interval $]0,+\infty[$. The
potential $\V$ is accordingly defined either
on~$\Rn$ or on $\D^\circ=\{q\in\Rn\;:\;q\cdot
  v_\alpha>0\;\forall\alpha\}$.

\bigskip

\strut\hbox{\bf Hypotheses 4.7 } \sl 
The vectors
$v_1,\ldots,v_N$ are nonzero and the cone
generated by them is proper.
The $f_\alpha$ are
$C^{m+1}$ ($m\ge2$) functions and
$$\eqalignno{&\sup f_\alpha=+\infty\,,\qquad
  \inf f_\alpha=0\,,
  &(4.13)\cr
  &f_\alpha^\prime(x)<0\quad\forall x\in
  \dom\,f_\alpha\,,
  &(4.14)\cr
  &f_\alpha^{(k)}(x)\cases{
     >0&if $k$ is even,\cr
     <0&if $k$ is odd,\cr}
                        \quad\forall x\ge a\,,
  &(4.15)\cr
  &f_\alpha^{(m+1)} \hbox{ is monotone on
           $[a,+\infty[$,}
  &(4.16)\cr}$$
where $a\ge0$ is a constant.  
Moreover, whichever of the three following
conditions i, ii, iii holds:

\smallskip

\item{i)} the vectors $v_\alpha$ verify 
$v_\alpha\cdot v_\beta\ge0$
$\forall\alpha,\beta$;

\smallskip

\item{ii)} all the functions
$f_\alpha$ are multiples of a single
function~$f$:
$$f_\alpha=c_\alpha f\,,\quad c_\alpha>0\,,
  \eqno(4.17)$$
such that 
$$x\mapsto x|f^\prime(x)| 
  \hbox{ is weakly  decreasing on }[a,+\infty[\,;
  \eqno(4.18)$$

\item{iii)} for all
$\alpha=1,\ldots,N$, the function
$f_\alpha$ is such that \rm 
$$\int_a^{+\infty}\mskip-14mu
  f_\alpha(x)\,dx<+\infty\,.
  \eqno(4.19)$$
\rm

\goodbreak
\bigskip

Now we are going to write down the statement of
global integrability. We skip for simplicity the
intermediate results concerning the mere
point~II of the Introduction.

\bigbreak

\strut\hbox{\bf Theorem 4.8 } \sl Assume
either one of the two following sets of
hypotheses: 

\item{a.} the ones of Proposition 4.3 plus
Hypotheses~4.5 and~4.6;  

\item{b.} Hypotheses 4.7 for a potential of the
form~(4.12).

\noindent
Then the Hamiltonian
system  
$$\dot p=-\nabla\V\,,\quad
  \dot q=p\,,
  \eqno(4.20)$$
is $C^m$-completely
integrable via asymptotic velocities on all of
the phase space. If moreover we are either in
Hypo\-theses~4.7, case iii, or the functions
$h_i$ of Hypo\-theses~4.1 and~4.3 verify
$\int_0^{+\infty}x^{i+1}h_i(x)\,dx<+\infty$ for
all $0\le i\le m+1$, then the asymptotic
map~$\A$ exists globally and it performs the
reduction to normal form.
\rm

\bigbreak

Here is what we can say about the
integrability of a slightly perturbed system. 

\bigbreak

\strut\hbox{\bf Theorem 4.9 } \sl Suppose
that we are in such hypotheses on~$\V$ that we
can apply Theorem~4.8. Let $K$ be a compact
set contained in the domain of~$\V$. Then there
exists an $\varepsilon>0$ with the following
property. Let $V$ be a $C^{m+1}$ real function
defined in~$\dom\V$, vanishing outside~$K$, and
such that 
$$\sup_K|\nabla V|\le\varepsilon\,.
  \eqno(4.21)$$
Then the thesis of Theorem~4.8 applies to
the system with potential $\V+V$. \rm

\goodbreak
\bigskip\bigskip

\centerline{{\bfuno 5. References}}

\bigskip

\frenchspacing
\parindent=30pt

\item{[A]} Arnold, V. I. (ed.) (1988). 
   {\bf Encyclopaedia of mathematical sciences 3,
     Dyna\-mical Systems III.} 
   Springer Verlag, Berlin.

\medbreak

\item{[Ga]} Galperin, G.A. (1982) {\it Asymptotic
   behaviour of particle motion under repulsive
   forces}. {\bf Comm. Math. Phys. 84},
   pp.~547--556.

\item{[Gu1]} Gutkin, E. (1985). 
   {\it Integrable  Hamiltonians with
   exponential potentials}. 
   {\bf Physica~D~16},
   pp.~398--404,
   North Holland, Amsterdam.

\medbreak

\item{[Gu2]} Gutkin, E. (1988).
  {\it Regularity of scattering trajectories in
  Classical Mechanics}.
  {\bf Comm. Math. Phys. 119},
  pp.~1--12.

\medbreak

\item{[GZ1]}  Gorni, G., \& Zampieri, G. (1989). 
  {\it Complete integrability for Hamiltonian
  systems with a cone potential}.  To appear in
  {\bf J. Diff. Equat.}

\medbreak

\item{[GZ2]}  Gorni, G., \& Zampieri, G. (1989). 
  {\it Reducing scattering
  problems under  cone potentials  to
  normal form by global canonical
  transformations}. 
  To appear in {\bf J. Diff. Equat.}

\medbreak

\item{[GZ3]}  Gorni, G., \& Zampieri, G.
  (1989).  {\it A class of integrable Hamiltonian
  systems including scattering of particles on
  the line with repulsive interactions.}
 To appear in {\bf  Differential and Integral Equations}.

\medbreak

\item{[He]} Herbst  (1974). {\it Classical
scattering  with long range forces}. \bf Comm.
Math. Phys. 35\rm, pp.~193--214.

\medbreak

\item{[Hu]} Hubacher A. (1989). {\it Classical
scattering theory in one dimension}. \bf Comm.
Math. Phys. 123\rm, pp.~353--375.

\medbreak

\item{[MNO]}  Moauro, V., Negrini, P., \& Oliva,
  W.M. (1989). {\it Analytic integrability for a
  class of cone potential mechanical systems}.
  In preparation.

\medbreak

\item{[M]} Moser, J. (1983). 
   {\it Various aspects  of integrable
   Hamiltonian systems}. 
   In {\bf Dynamical Systems}
   (C.I.M.E. Lectures, Bressanone 1978),
   pp.~233--290, sec. print.,
   Birkh\"auser, Boston.

\medbreak

\item{[OC]} Oliva, W.M., 
   \& Castilla M.S.A.C. (1988).
   {\it On a class of $C^\infty$-integrable 
   Hamiltonian systems}. 
   To appear  in \bf Proc. Royal Society
   Edinburgh\rm.

\medbreak

\item{[S]} Simon B. (1971). {\it Wave operators
 for classical particle scattering}.
\bf Comm. Math. Phys. 23\rm, pp.~37--48.

\vfill

\hbox to 3cm{\hrulefill}

\smallskip

\noindent
This research was made under the auspices of the
C.N.R. (Italian National Research Council) and of
the M.U.R.S.T. (Italian Ministery for University
and Scientific and Technological Research).

\bye